\documentclass[twocolumn,showpacs,preprintnumbers,amsmath,amssymb,superscriptaddress]{revtex4-2}
\usepackage{graphicx}
\usepackage{dcolumn}
\usepackage{bm}
\usepackage{lmodern}
\usepackage{epsfig}
\usepackage{multirow}
\usepackage{longtable}
\usepackage{graphics}
\usepackage{graphicx}
\usepackage{subfigure}
\usepackage{afterpage}
\usepackage{enumerate}
\usepackage[T1]{fontenc}

\usepackage{natbib}
\usepackage[utf8]{inputenc}

\begin{document}

\title{Exploring the competition between $\alpha$-decay and proton radioactivity: A comparative study of proximity potential formalisms}

\author{A. Adel}
\email{ahmedadel@sci.cu.edu.eg}
\affiliation{Physics Department, Faculty of Science, Cairo
University, 12613 Giza, Egypt}

\author{Karim H. Mahmoud}
\affiliation{Physics Department, Faculty of Science, Cairo
University, 12613 Giza, Egypt}

\author{Haitham A. Taha}%
\affiliation{Department of Physics, Zewail City for Science and Technology, Egypt}
\date{\today}


\begin{abstract}

We have conducted a comprehensive and systematic study of the proton radioactivity and $\alpha$-decay half-lives of neutron-deficient nuclei. This investigation involved the utilization of various Proximity potentials and also considered the incorporation of thermal effects. For the half-life calculations, we employed both temperature-independent and temperature-dependent interaction potentials. We observed that proton radioactivity serves as the dominant mode of decay for nuclides situated very close to the proton drip-line. We explored a universal curve that examines the correlation between the decimal logarithm of experimental half-lives and the negative decimal logarithm of the penetrability for both proton radioactivity and $\alpha$-decay.

\end{abstract}

\maketitle

\section{\label{sec:level1}Introduction}
The investigation of exotic nuclei containing an extreme number of nucleons has consistently garnered considerable interest, with a primary focus on $\alpha$-decay and proton radioactivity 
\cite{Cheng2020,adel2021proton,Qian2016,Basu2005,Chen2019,Chen2019a,Delion2006,Adel2018,Ismail2020}. 
$\alpha$-decay offers valuable insights into nuclear structure properties, providing abundant information on shell and pairing effects \cite{Ismail2010}, neutron skin thickness \cite{Seif2017}, isospin asymmetry \cite{Seif2011}, nuclear deformation \cite{Ismail2014}, and shape staggering identification \cite{Andreyev1999}. Moreover, $\alpha$-decay plays a crucial role in identifying recently synthesized superheavy nuclei \cite{Oganessian2015}. On the other hand, the emission of protons from the nuclear ground state imposes restrictions on the formation of increasingly exotic nuclei on the proton-rich side of the beta stability valley \cite{Buck1992}. One proton radioactivity can serve as a valuable tool for obtaining spectroscopic information since the decaying proton is an unpaired proton that does not fully occupy its orbit. The rates of these decays are highly sensitive to the $Q$ values and orbital angular momenta, which, in turn, assist in determining the orbital angular momenta of the emitted protons \cite{Basu2005}.

Driplines represent the boundaries of the nuclear landscape, defined by nuclear binding energies. The proton drip-line marks the border where nuclei become unbound to proton emission from their ground states \cite{Blank2008}. Beyond the proton drip-lines, the proton separation energy becomes negative, indicating insufficient binding energy to retain protons within the nucleus. These proton-rich nuclei have positive $Q$-values for proton emissions, which drives the spontaneous tendency to shed excess protons. This region of proton emitters presents a significant challenge for nuclear theory in predicting the properties of unstable states.

Proton radioactivity was experimentally confirmed in 1970, with evidence of proton emission from the high-spin isomeric state of $^{53}\textrm{Co}^m$ to the ground state of $^{52}\textrm{Fe}$ [22]. Subsequently, proton emissions from various ground states (e.g., $^{151}\textrm{Lu}$ and $^{147}\textrm{Tm}$) and isomeric states have been discovered, with a combined total of about 28 proton emitters decaying from ground states and 20 from isomeric states \cite{Blank2008}. With the advancement of new experimental facilities and radioactive beams, more new proton emitters are anticipated to be observed, shedding light on the detailed nuclear structure in the proton dripline region.

Proton radioactivity can be effectively addressed using the Wentzel-Kramers-Brillouin (WKB) method, as this process can be treated as a quantum tunneling effect through a potential barrier, similar to $\alpha$-decay \cite{Deng2019}. Several methods have been employed to investigate proton radioactivity, including the density-dependent M3Y effective interaction \cite{Bhattacharya2007,Basu2005}, the JLM interaction \cite{Bhattacharya2007}, the unified fission model \cite{JianMin2010}, the generalized liquid drop model (GLDM) \cite{Dong2009}, the cluster model \cite{HongFei2009}, the deformed density-dependent model \cite{Qian2016}, the Gamow-like model \cite{Zdeb2016}, the Coulomb and proximity potential model for deformed nuclei (CPPMDN) \cite{Santhosh2017}, the covariant density functional (CDF) theory \cite{zhao2014proton}, analytic formulas \cite{Zhang2018}, the distorted-wave Born approximation (DWBA) \cite{Aaberg1997}, the two-potential approach (TPA) \cite{Aaberg1997}, the quasiclassical method \cite{Aaberg1997}, and others.

Several theoretical approaches have been developed for describing $\alpha$-decay, including the density-dependent cluster model \cite{Xu2005,adel2015fine,Ismail2014a}, the generalized liquid-drop model (GLDM) \cite{Zhang2007}, the modified generalized liquid-drop model (MGLDM) \cite{Santhosh2018b,Akrawy2019}, the effective liquid-drop model (ELDM) \cite{Goncalves1993,Cui2018}, the coupled channel approach \cite{Delion2006a,Peltonen2007}, the fission-like model \cite{Poenaru1979}, and the Coulomb and Proximity potential model (CPPM) \cite{Santhosh2017a,Santhosh2018}. In addition, various analytical and empirical formulas have been proposed to predict $\alpha$-decay half-lives \cite{Viola1966,Sobiczewski1989,Budaca2016,Poenaru2007,Akrawy2018}. The key element in most of these studies is the adopted model for the nuclear part of the potential.

Blocki \textit{et al.} introduced the proximity potential as a means to address heavy ion reactions \cite{blocki1977proximity}. As a nucleus-nucleus interaction potential, it relies on the proximity force theorem \cite{blocki1977proximity,Blocki1981}, represented as the product of a factor dependent on the mean curvature of the interaction surface and a universal function (dependent on the separation distance) that remains unaffected by the masses of the colliding nuclei \cite{dutt2010systematic}. One of its key advantages lies in its straightforward and accurate formalism. However, to overcome the limitations of the original version of the proximity potential (Prox.1977) \cite{blocki1977proximity}, several improvements and modifications have been proposed. These developments have involved either enhancing the form of the surface energy coefficients \cite{Krappe1979,Moller1995,Pomorski2003} or introducing an improved universal function or another parameterization for nuclear radius \cite{Blocki1981,dutt2010systematic,bass1973threshold,bass1974fusion}.

Several comparative studies have been conducted by different researchers, including Yao \textit{et al.} \cite{Yao2015}, Ghodsi \textit{et al.} \cite{ghodsi2016systematic}, and Santhosh \textit{et al.} \cite{Santhosh2017b}, to explore various proximity potential formalisms used in describing $\alpha$-decay and heavy particle radioactivity. Furthermore, these proximity potential formalisms have found application in investigating heavy ion fusion \cite{Santhosh2009,Santhosh2014} and ternary fission \cite{Santhosh2014a,Santhosh2015}.

The interplay between $\alpha$-decay and proton emission becomes evident in the region of neutron-deficient nuclei \cite{Wang2017}. Prompted by this observation, we have been motivated to thoroughly investigate the competition between these two decay modes, aiming to ascertain which one holds dominance. This investigation provides a valuable testing ground for theoretical models. Thus, the main objective of this research is to examine the appropriateness of various proximity potentials concerning the interplay between $\alpha$-decay and proton radioactivity in neutron-deficient nuclei. Moreover, we investigate the impact of thermal effects by examining temperature-independent and temperature-dependent interaction potentials for half-life calculations. Subsequently, we compare these calculated values with the available experimental data.

\section{\label{sec:level2}Theoretical Framework}
\subsection{Half-life time calculation}

To determine the half-life time of radioactive nuclei, it is necessary to calculate the total interaction potential $V(r)$ \cite{adel2015fine,Deng2019}. It can be mathematically represented as follows:
 \begin{equation}
    V(r) = V_N(r) + V_C(r) + V_{\ell}(r),
\end{equation}
where $V_C(r)$,$V_N(r)$, and $V_{\ell}(r)$ are the coulomb, nuclear, and centrifugal potentials, respectively. In this study, we employ the proximity potential formalism to compute the nuclear potential $V_N(r)$. The Coulomb potential, $V_C(r)$, is postulated as the potential arising from a uniformly charged sphere with a radius of $R$.  It can be expressed as follows:

\begin{equation}
V_C(r)= \begin{cases}\frac{Z_1 Z_2 e^2}{2 R}\left[3-\left(\frac{r}{R}\right)^2\right], & r \leq R, \\ \frac{Z_1 Z_2 e^2}{r}, & r>R,\end{cases}
\end{equation}
where $R = R_1 + R_2$ is given by the sum of the radii of daughter nucleus $R_1$ and emitted particle (alpha, proton, or cluster) $R_2$. While $Z_1$ and $Z_2$ are the proton numbers of the daughter nucleus and emitted particle, respectively.

We utilize the Langer modified form for the centrifugal potential $V_{\ell}(r)$ due to the necessity of the correction term $\ell(\ell+1) \rightarrow (\ell+\frac{1}{2})^2$ for one-dimensional problems \cite{Langer1937}. This form can be expressed as follows:
\begin{equation}
V_{\ell}(r) = \frac{\hbar^2(\ell+\frac{1}{2})^2}{2\mu r^2},
\end{equation}
where $\ell$ is the angular momentum carried by the emitted particle. We can determine the minimum angular momentum $\ell_{min}$ using spin and parity conservation rules, and $\mu$ is the reduced mass of the emitted particle and daughter nucleus where $\mu = \frac{m_1m_2}{m_1+m_2}$, where $m_1$ and $m_2$ are the masses of the daughter nucleus and emitted particle, respectively.


Once the total interaction has been computed, the penetration probability of the particle can be determined using the following expression:

\begin{equation}
P = \exp\left(-2\int_{R_{in}}^{R_{out}} k(r) \,dr\right),
\end{equation}
Here, the classical turning points $R_{in}$ and $R_{out}$ are determined from $V(R_{in})=V(R_{out})=Q$. The wave number $k(r)$ of the emitted particle given as $k(r) = \sqrt{\frac{2\mu}{\hbar^2}|V(r)-Q|}$ and $Q$ is the decay energy.

The knocking frequency (or assault frequency) $\nu$ is the number of assaults on the barrier per second and it is related to oscillation frequency $\omega$. It is given by \cite{ghodsi2016systematic}:
\begin{equation}
\nu = \frac{\omega}{2\pi} = \frac{2E_{\nu}}{h}
\end{equation}
where $E_{\nu}$ is zero-point vibration energy, which, in the case of a proton  it is related to its $Q$-value ($Q_p$) \cite{poenaru1986odd} as follows:
\begin{equation}
E_\nu= \begin{cases}0.1045 \, Q_p, & \text { for even-} Z \text {-even-} N \text {-parent nuclei, } \\ 0.0962 \, Q_p, & \text { for odd-} Z \text {-even-} N \text {-parent nuclei, } \\ 0.0907 \, Q_p, & \text { for even-} Z \text {-odd-} N \text {-parent nuclei, } \\ 0.0767 \, Q_p, & \text { for odd-} Z \text {-odd-} N \text {-parent nuclei. }\end{cases}
\end{equation}
%
%
%
In the case of an emitted alpha particle or cluster, the value of the empirical zero-point vibration energy $E_{\nu}$ can be obtained from the following expression \cite{Poenaru1985,Kumar2021}:
$$\begin{aligned} E_{v}{=}Q\left\{ 0.056+0.039\;\mathrm{exp}\left[ \frac{4-A_{2}}{2.5} \right] \right\} ,\hbox { for }A_{2} \ge 4.\nonumber \\ \end{aligned}$$

The half-life time is subsequently calculated using the formula:
\begin{equation}
T_{\frac{1}{2}} = \frac{\ln 2}{\lambda}
\end{equation}
where $\lambda$ is the decay constant given by $\lambda = \nu P$.

\subsection{The proximity potential}
For evaluating the nuclear potential term in the total interaction between the daughter nucleus and the emitted particle, we will employ six distinct proximity potentials: CW76 \cite{christensen1976evidence}, Denisov \cite{denisov2002interaction}, Guo2013 \cite{guo2013study}, Prox00 DP \cite{dutt2010systematic}, Bass73 \cite{bass1973threshold,bass1974fusion}, and Denisov DP \cite{dutt2010systematic}.

\subsubsection{The proximity potential Guo2013}

Guo\textit{ et al.} introduced a novel universal function of proximity potential model labeled as Guo2013 by employing the double folding model with density-dependent nucleon-nucleon interaction and fitting the universal functions of numerous reaction systems \cite{guo2013study}. The expression for this proximity potential is as follows:
%
%
\begin{equation}
V_N(r) = 4\pi\gamma b\frac{R_1R_2}{R_1+R_2}\phi(\xi),
\end{equation}
In this expression, the parameter $b$ represents the diffuseness of the nuclear surface and is assigned a value of unity and $\gamma$ is the surface coefficient and is obtained by the following expression:
\begin{equation}
\gamma = 0.9517\left[1-1.7826\left(\frac{N-Z}{A}\right)^2\right]
\end{equation}
In this context, $N$, $Z$, and $A$ denote the neutron, proton, and mass numbers, respectively, of the parent nucleus. The values of $R_1$ and $R_2$ can be expressed as follows:
\begin{equation}
\begin{array}{cc}
R_i = 1.28A_i^{\frac{1}{3}} - 0.76 + 0.8A_i^{-\frac{1}{3}} & (i = 1,2).
\end{array}
\end{equation}
The universal function $\phi(\xi)$ is given by the following expression:
\begin{equation}
\phi(\xi) = \frac{p_1}{1+exp(\frac{\xi+p_2}{p_3})}
\end{equation}
where $p_1$, $p_2$ and $p_3$ are adjustable parameters whose values are -17.72, 1.30 and 0.854, respectively and $\xi$ is given by $\xi = \frac{r-R_1-R_2}{b}$.\\
\subsubsection{The proximity potential CW76}
Christensen and Winther analyzed the heavy-ion elastic scattering data and came up with an empirical nuclear potential\cite{christensen1976evidence} which is expressed as follows:
\begin{equation}
V_N(r) = -50\frac{R_1R_2}{R_1+R_2}\phi(r-R_1 - R_2)
\end{equation}
where $R_1$ and $R_2$ are given by:
\begin{equation}
\begin{array}{cc}
R_i = 1.233A_i^{\frac{1}{3}} - 0.978A_i^{-\frac{1}{3}} & (i = 1,2)
\end{array}
\end{equation}
and the universal function is given by $\phi(r - R_1 - R_2) = exp(-\frac{r- R_1 - R_2}{0.63})$.
\subsubsection{The proximity potential Bass73}
Bass developed a nuclear potential expression in terms of the difference between finite and infinite separation $\xi$ based on the liquid drop model\cite{bass1973threshold,bass1974fusion}. The expression is given as follows:
\begin{equation}
\begin{split}
V_N(r) & = -4\pi\gamma\frac{dR_1R_2}{R}exp\left(-\frac{\xi}{d}\right) \\
    & = \frac{-da_sA_1^{\frac{1}{3}}A_2^{\frac{1}{3}}}{R}exp\left(-\frac{r-R}{d}\right)
\end{split}
\end{equation}
$\gamma$ represents the specific surface energy of the liquid drop model, while $d$ is the range parameter whose value is 1.35 fm. $a_s$ is the surface term of the Bethe Weizsacker semi-empirical mass formula, whose value is 17.0 MeV. R is given by the sum of the half-maximum density radii $R = R_1 + R_2 = r_0\left(A_1^{\frac{1}{3}}+A_2^{\frac{1}{3}}\right)$, where $r_0 = 1.07$, $R_1$, $A_1$. $R_2$ and $A_2$ are the radii and mass numbers of the daughter nucleus and emitted particle respectively.

\subsubsection{The proximity potential Denisov}
A nuclear potential expression was presented by Denisov, where he worked with 119 spherical or quasi-spherical even-even nuclei around the $\beta$ stability line in the semi-microscopic approximation between all possible nucleus-nucleus combinations\cite{denisov2002interaction}. The nuclear potential is given by:
\begin{equation}
\begin{split}
V_N(r) = & -1.989843\frac{R_1R_2}{R_1+R_2}\phi(\xi)\left[1 + 0.003525139 \right. \\
 & \times \left(\frac{A_1}{A_2} + \frac{A_2}{A_1}\right)^{\frac{3}{2}}- 0.4113263(I_1 + I_2)\left. \right]
\end{split}
\end{equation}
$I_1$ and $I_2$ represent the isospin asymmetry of the daughter and emitted particle which is given by $I_i = \frac{N_i - Z_i}{N_i + Z_i}$ $(i=1,2)$. $Z_1$, $N_1$, $R_1$ and $Z_2$, $N_2$, $R_2$ are the proton number, neutron number, and the effective nucleus radii of the daughter and emitted particle respectively. The effective nucleus can be obtained via:

\begin{equation}
\begin{split}
R_i & = R_{ipar}\left(1-\frac{3.413817}{R_{ipar}^2}\right) \\
& + 1.284589\left(I_i - \frac{0.4A_i}{A_i + 200}\right)      (i=1,2)
\end{split}
\end{equation}
where the radius of the emitted particle $R_{ipar}$ is given by:
\begin{equation}
R_{ipar} = 1.240A_i^{\frac{1}{3}}\left(1 + \frac{1.646}{A_i} - 0.191I_i\right)
\end{equation} \\
The universal function $phi(\xi)$ is given by:
\begin{equation}
\phi(\xi) =
\left\{
\begin{array}{cc}
\begin{split}
   &\hspace{-0.2cm}\left[1-\xi^2\left[0.05410106\frac{R_1R_2}{R_1 + R_2}\exp\left(-\frac{\xi}{1.760580}\right)\right]\right.\\
   & -0.5395420(I_1 + I_2)\exp\left(-\frac{\xi}{2.424408}\right)\bigg] \\
   & \times \exp\left(-\frac{\xi}{0.7881663}\right) \quad \xi \geq 0,\\
   & 1 - \frac{\xi}{0.7881663} + 1.229218\xi^2 - 0.2234277\xi^3\\
   & -0.1038769\xi^4 - \frac{R_1R_2}{R_1 + R_2}(0.1844935\xi^2\\
   & + 0.07570101\xi^3) + (I_1 + I_2)(0.04470645\xi^2\\
   & + 0.03346870\xi^3) \quad -5.65 \leq \xi \leq 0
\end{split}
\end{array}
\right\}
\end{equation}
where $\xi = r - R_1 - R_2 - 2.65$.

\subsubsection{The proximity potential gp77}
This proximity potential was used by Santhosh et al.\cite{santhosh2013theoretical} and it is given as follows:
\begin{equation}
V_N(r) = 4\pi\gamma b\frac{C_1C_2}{C_1+C_2}\phi(\xi)
\end{equation}
which is similar to the Guo2013 potential but instead of using the radii of the daughter and emitted nuclei, we use the S$\ddot{\text{u}}$ssman central radii of the fragments, which are related as follows:
\begin{equation}
C_i = R_i - \frac{b^2}{R_i}	
\end{equation}
where $i = 1,2$ representing daughter and emitted particle respectively. The radii $R_i$ are the same as in Guo2013. In this potential, the nuclear surface tension coefficient is given as follows:
\begin{equation}
\gamma = \gamma_0\left[1 - k_s\left(\frac{N-Z}{A}\right)\right]
\end{equation}
where Z, N, and A are the proton number, neutron number, and mass number of the parent nucleus respectively, and the coefficients $\gamma_0$ and $k_s$, in this case, are given as $\gamma_0 = 0.9517$ and $k_s = 1.7826$. The universal function $\phi(\xi)$ is given as:
\begin{equation}
\phi(\xi) =
\left\{
\begin{array}{cc}
\begin{split}
   &-1.7817 + 0.927\xi + 0.0169\xi^2 - 0.05148\xi^3\\
   &\quad \xi \leq 1.9475,\\
   &-4.41exp(-\xi/0.7176)  \quad  \xi \geq 1.9475.
\end{split}
\end{array}
\right\}
\end{equation}

\subsection{Temperature dependent proximity potential}
In the previous sections, we worked with nuclei at the ground state(no temperature dependence). In this section, thermal effects are taken into account, where temperature-dependent forms of $R$, $\gamma$, and $b$ are presented\cite{gharaei2016temperature}. They are expressed as follows:
\begin{equation}
\begin{array}{cc}
    R_i(T) = R_i(T= 0)[1 + 0.0005T^2]fm & (i= 1,2)
\end{array}
\end{equation}
\begin{equation}
\gamma(T) = \gamma(T= 0)\left[1 - \frac{T - T_b}{T_b}\right]^{\frac{3}{2}}
\end{equation}
\begin{equation}
b(T) = b(T = 0)\left[1 + 0.009T^2\right]
\end{equation}
$T_b$ represents the temperature associated with near coulomb barrier energies. In this work b(T = 0) = 1 and $R_i(T= 0)$ is the normal radius formula associated with each proximity potential we have worked with. Instead of the form written in equation $(31)$ we will use a different form of $\gamma(T)$ given as $\gamma(T) = \gamma(0)(1 - 0.07T)^2$\cite{ghorbani2020temperature}. We can determine the temperature $T$ (in Mev) using the following expression:
\begin{equation}
E^* = E_{kin} + Q_{in} = \frac{1}{9}AT^2 - T
\end{equation}
$E^*$ represents the excitation energy of the parent nucleus, while $A$ is its mass number. $Q_{in}$ is the entrance channel $Q$ - value of the system. To obtain $E_{kin}$, which is the kinetic energy of the emitted particle we can use the following expression:

\begin{equation}
E_{kin} = \left(\frac{A_d}{A_p}\right)Q
\end{equation}

\section{\label{sec:level3}RESULTS AND DISCUSSION}

\begin{figure*}[htp!]
\centering
\subfigure[\label{Fig1a}]{\includegraphics[width=8.6cm]{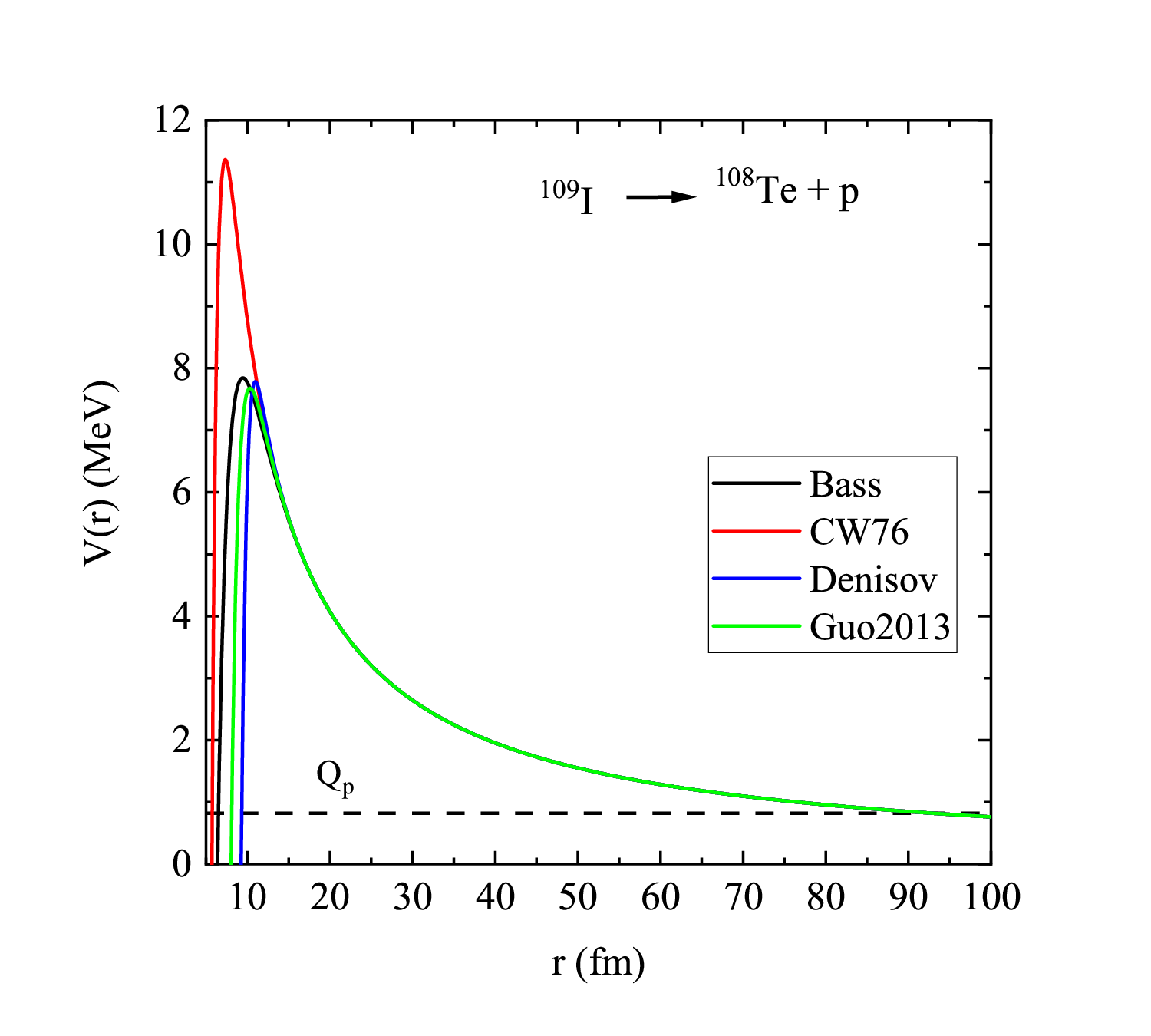}}
 \subfigure[\label{Fig1b}]{\includegraphics[width=8.6cm]{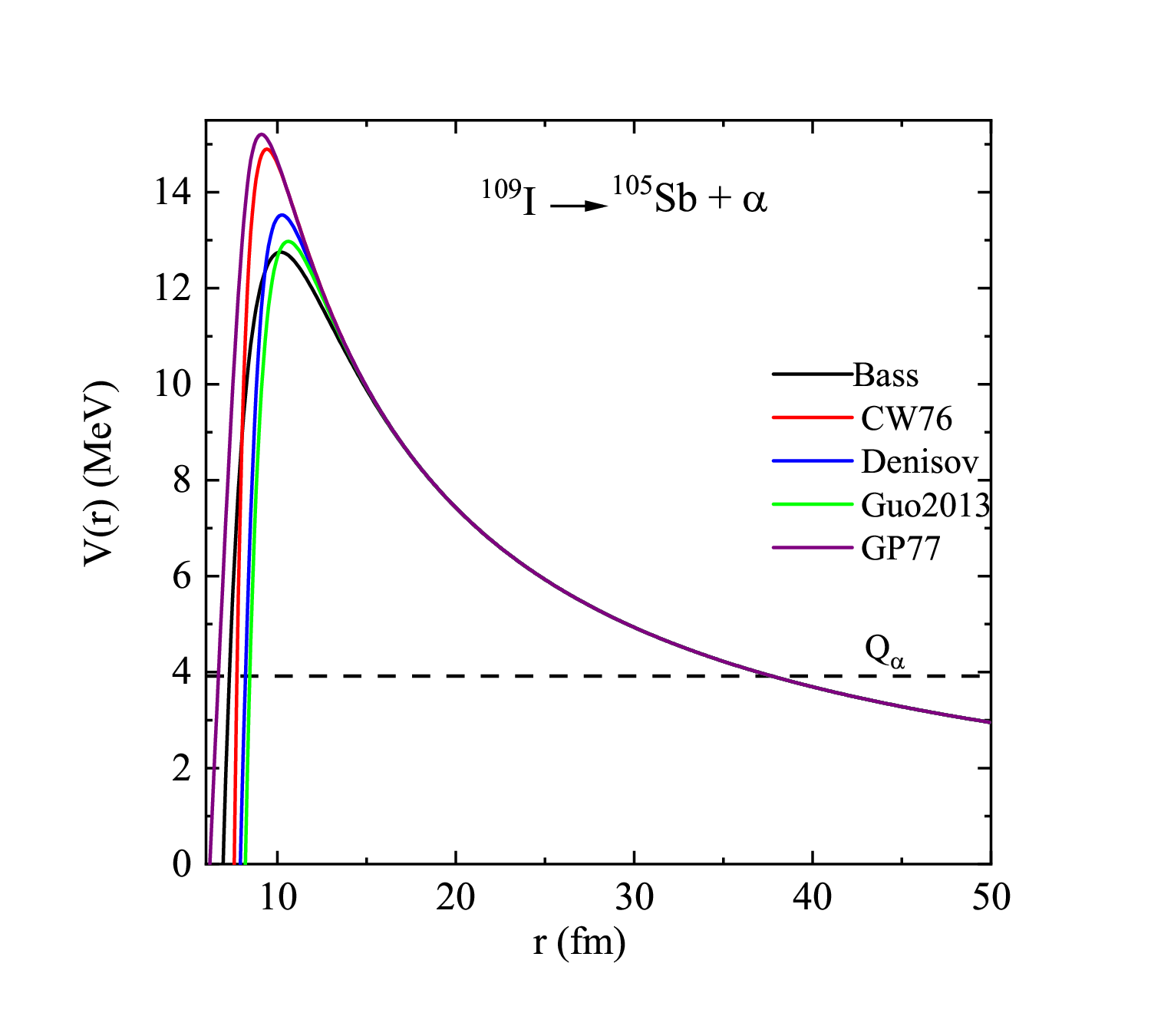}}

 \caption{\label{Fig1} The distributions of total emitted proton-core interaction potential $V (r)$ including the four versions of proximity potentials in case of proton emission and five versions of proximity potentials in case of $\alpha$ decay for the decay of $^{109}\textrm{I}$.}
\end{figure*}

We conducted a comprehensive investigation of the half-lives of numerous neutron-deficient nuclei, spanning the range of atomic numbers from $Z=53$ to 81. These nuclei decay through  proton emission and/or $\alpha$-decay. Our analysis utilized five distinct proximity potentials: Bass73, CW76, Guo2013, Denisov, and gp77. In Figure~\ref{Fig1}, the total potential $V(r)$ for the decay of neutron-deficient $^{109}\textrm{I}$ is depicted, utilizing different proximity potentials. Figure~\ref{Fig1}(a) showcases the proton decay, while Fig.~\ref{Fig1}(b) provides insight into the $\alpha$-decay process. 
The dashed line represents the $Q$-value for each decay. In Fig~\ref{Fig1}(a), the Coulomb barrier height for proton decay is shown to be highest among various proximity potentials, with CW76 potential displaying the maximum barrier. Conversely, for $\alpha$-decay, both GP77 and CW76 exhibit the highest Coulomb barriers. The detailed results of the half-life calculations are presented in both Tables \ref{tab:table1} and~\ref{tab:table2}. In Tables~\ref{tab:table1} and~ \ref{tab:table2}, the first column provides information about the parent nucleus. Additionally, the symbol "$m$" denotes the first isomeric state for the parent nucleus. Moving to the second and third columns in both Tables~\ref{tab:table1} and~ \ref{tab:table2}, we present the experimental decay energy, denoted as $Q_{p/\alpha}$, and the transferred angular momentum $\ell_\textrm{min}$, respectively. Subsequent columns in these tables display the logarithmic half-lives calculated using the aforementioned potential models. Incorporating considerations for temperature effects, we conducted a comparison between the temperature-independent half-lives labeled as "T-IND" and the temperature-dependent half-lives labeled as "T-DEP" in both Tables~\ref{tab:table1} and~\ref{tab:table2}. Finally, the last column contains the available experimental half-life data for the decaying nucleus.

%
%

It is noteworthy that our theoretical calculations for both proton decay and $\alpha$-decay half-lives across different parent nuclei exhibit remarkable consistency and robust agreement with the available experimental data. Any discrepancies between the calculated and experimental half-lives may be attributed to the uncertainties inherent in the measurements of experimental $Q$ values, which have a notable impact on the outcomes. This underscores the effectiveness of our calculations, resulting in a small standard deviation ($\sigma$) between the computed and experimental logarithmic half-lives. In essence, these findings affirm the robustness and reliability of our understanding of these fundamental processes. We have calculated the standard deviation, represented as $\sigma$, between the experimental and computed values of the logarithmic half-lives as:
\begin{equation}
   \sigma = \left[\frac{1}{n-1}\sum_{i=1}^n(\log_{10}T_{1/2}^{calc.} - \log_{10}T_{1/2}^{expt.})^2\right]^{1/2}
\end{equation}
The standard deviation values ($\sigma$) for different potentials are as follows: 1.3229, 1.9437, 0.288, and 0.8165 for Bass73, CW76, Guo2013, and Denisov, respectively, in the case of proton emission. For $\alpha$-decay, the $\sigma$ values are: 1, 1.25, 2.5249, 2.5981, and 0.4675 for Bass, CW76, Denisov, Guo2013, and gp77, respectively. These findings highlight that among the potentials used for $\alpha$-decay half-life calculations, gp77 stands out as the most effective, while Guo2013 appears to be the most suitable for proton emission. 

\begin{figure*}[htp!]
\centering
\subfigure[\label{Fig2a}]{\includegraphics[width=8.6cm]{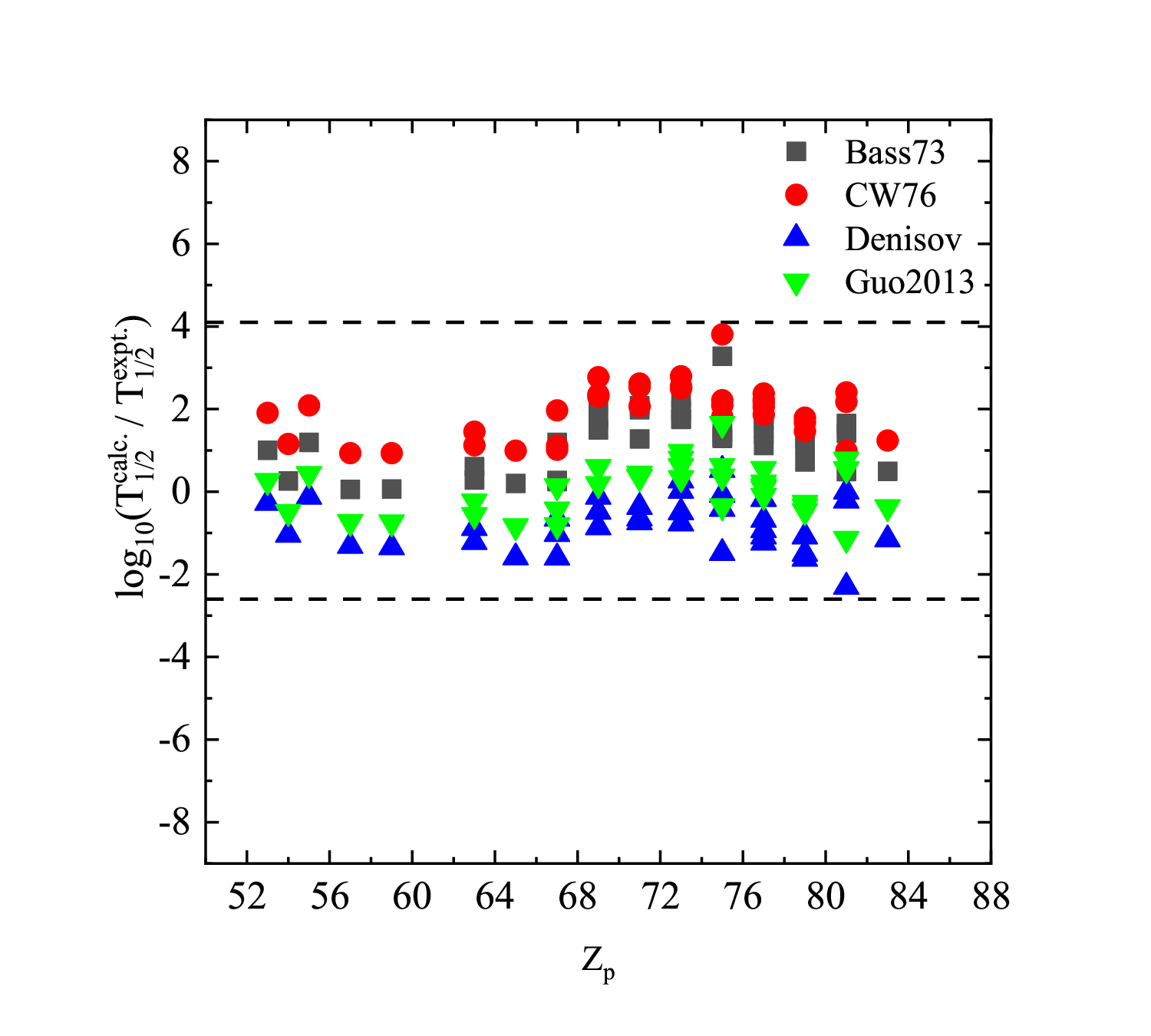}}
 \subfigure[\label{Fig2b}]{\includegraphics[width=8.6cm]{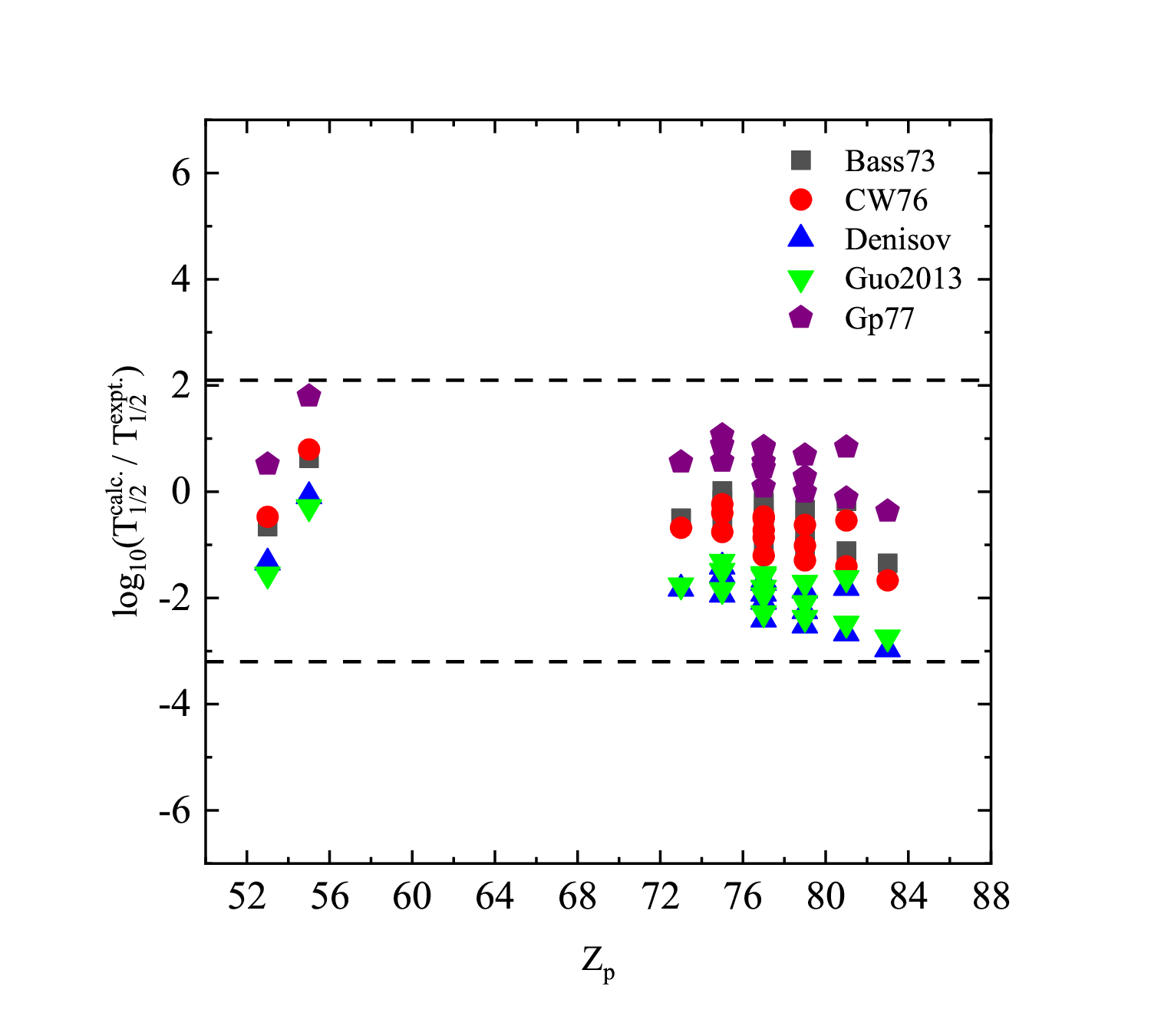}}

 \caption{\label{Fig2} (a)Deviation of the calculated proton emission half-lives, $T_{1/2}^\textrm{Bass73}$, $T_{1/2}^\textrm{CW76}$, $T_{1/2}^\textrm{Guo2013}$, and $T_{1/2}^\textrm{Denisov}$ with the corresponding experimental data for the neutron-deficient proton emitters listed in Tables \ref{tab:table1} and \ref{tab:table2} (b) Deviation of the calculated $\alpha$ emission half-lives,  with the corresponding experimental data for the neutron-deficient $\alpha$ emitters listed in Tables \ref{tab:table1} and \ref{tab:table2}.}
\end{figure*}

In Fig.~\ref{Fig2}(a), we observe the discrepancies in proton decay half-lives among various proton emitters, as listed in Table~\ref{tab:table1}, in relation to their respective experimental values. These deviations, $\log_{10}(T_{1/2}^\textrm{calc.}/T_{1/2}^\textrm{expt.})$, are graphed as a function of the proton number ($Z_p$) of the parent nucleus. Notably, Figure~\ref{Fig2}(a) distinctly illustrates that the calculated proton decay half-lives exhibit deviations from experimental data, typically within an order of magnitude of 2, across multiple potential scenarios. Our analysis reveals that the majority of data points cluster around $\log_{10}(T_{1/2}^\textrm{calc.}/T_{1/2}^\textrm{expt.}) = 0$, indicating that most calculated proton decay half-lives align favorably with experimental data for different proton emitters. In Fig.~\ref{Fig2}(b), we illustrate the deviations between the computed and experimental $\alpha$-decay half-lives. It's apparent from the graph that a significant portion of the data points is positioned near $\log_{10}(T_{\text{calc}}/T_{\text{expt}}) = 0$, suggesting a notable alignment between the majority of calculated $\alpha$-decay half-lives and their experimental counterparts. The conclusions drawn from the Figs.~\ref{Fig2}(a) and \ref{Fig2}(b) are quite clear. They emphasize that, based on the data presented, gp77 stands out as the most effective potential for $\alpha$-decay half-life calculations, whereas Guo2013 seems to be the most suitable choice for proton emission calculations.

\begin{figure*}[htp!]
\centering
\subfigure[\label{Fig3a}]{\includegraphics[width=8.6cm]{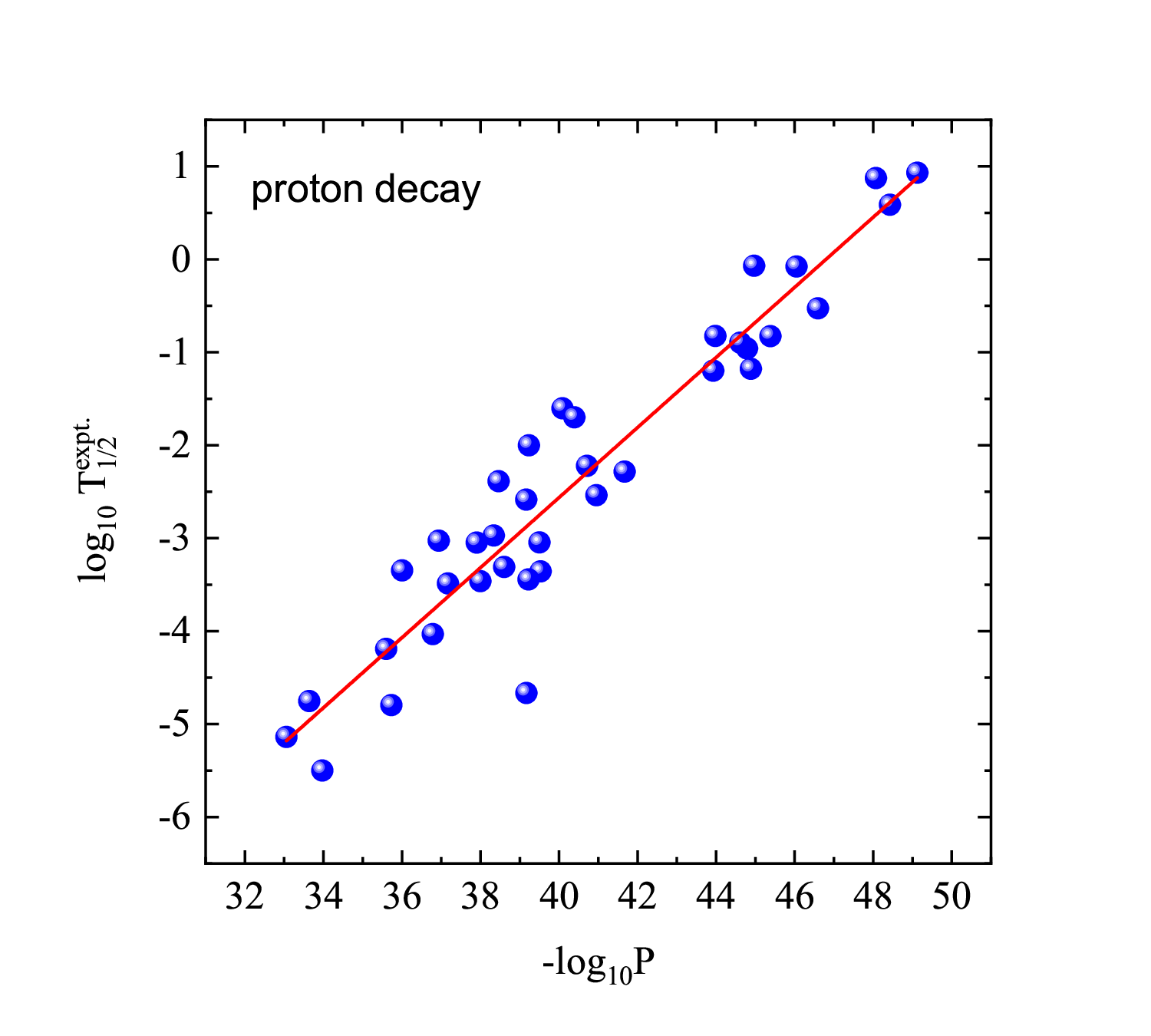}}
 \subfigure[\label{Fig3b}]{\includegraphics[width=8.6cm]{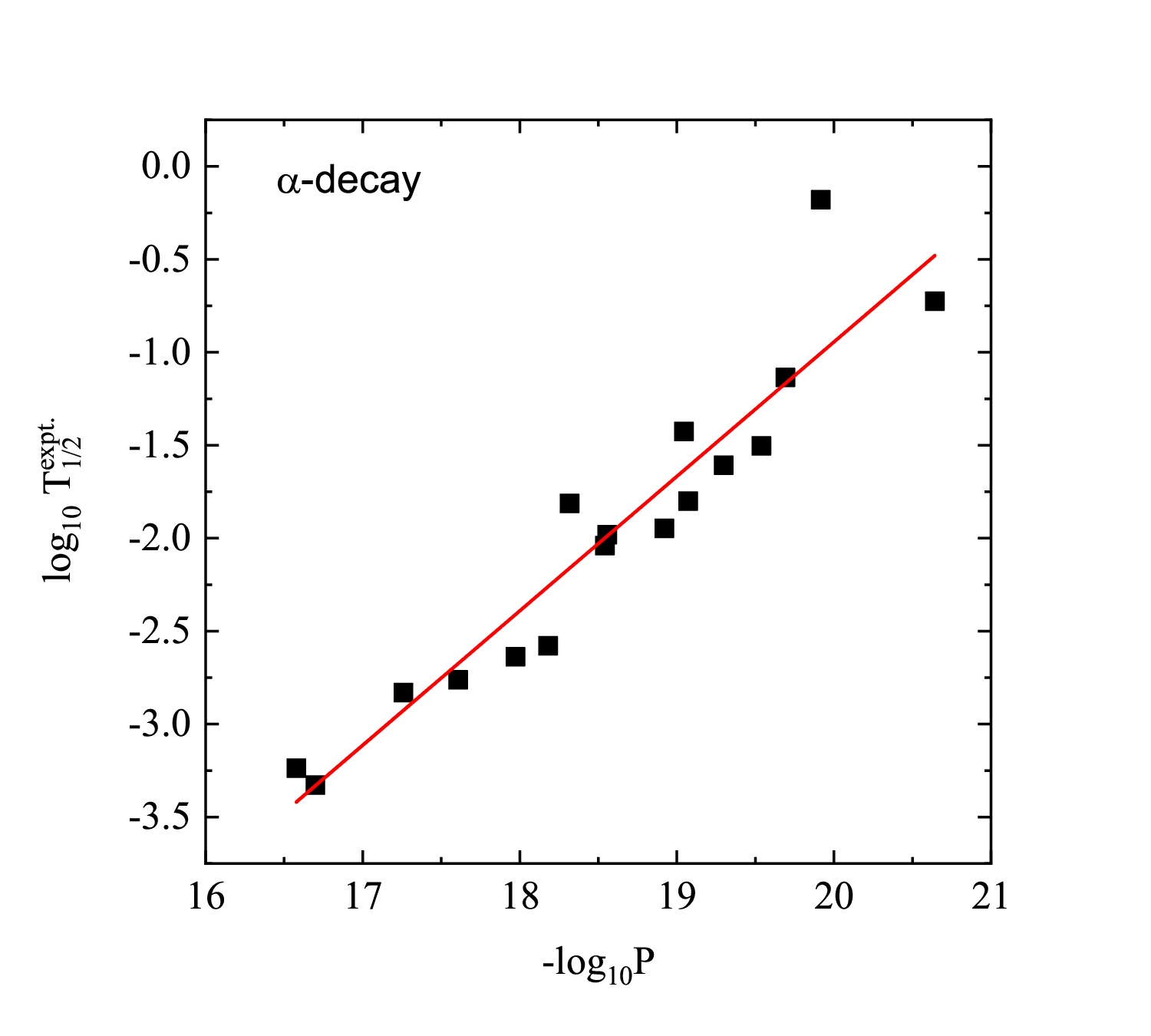}}

 \caption{\label{Fig3} A universal curve based on the WKB penetration probability, $P$, using the Guo2013  potential for proton emission and gp77 for $\alpha$ emission. The logarithm of the experimental half-life times are plotted versus the $-\log_{10}P$ (a) for proton decay (b) for $\alpha$-decay listed in Tables \ref{tab:table1} and \ref{tab:table2}. The fitted solid line for each case are shown in the corresponding figure.}
\end{figure*}

The universality of the $\alpha$- and cluster decay phenomena can be effectively demonstrated by creating a graphical representation that plots the logarithm of the experimental half-life against the negative logarithm of the penetration probability through the barrier, as previously explored in Refs.~ \cite{Poenaru1991,Poenaru2006,Poenaru2011,Ismail2015}. It is intriguing to investigate whether this correlation can also be applied to proton radioactivity of neutron-deficient nuclei as shown in Fig.~\ref{Fig3}. Figure ~\ref{Fig3}(a) pertains to proton decay, whereas Fig.~\ref{Fig3}(b) focuses on $\alpha$-decay. These figures visually depict the relationship between the decimal logarithm of experimental half-lives ($\log_{10} T_{1/2}^\textrm{expt.}$) and the negative decimal logarithm of penetrability ($-\log_{10} P$). The penetrability values are computed theoretically within the framework of the WKB approximation, employing the Guo2013 potential for proton emission and gp77 for $\alpha$-emission. The straight-line relationship evident in these plots affirms the robustness of our calculations and underscores the credibility of our methods.

Qi \textit{et al.} \cite{qi2009universal} have introduced a linear universal decay formula (UDL) that exhibits remarkable accuracy in predicting half-lives across various cluster types and isotopic series, particularly for ground-state to ground-state transitions. This formula offers valuable physical insights and is recognized for its universal applicability. By employing a similar analytical approach as Qi \textit{et al.} \cite{Qi2012} and accounting for the centrifugal barrier's impact, we can discern a distinct and coherent graphical pattern in the proton decay half-lives \cite{adel2021proton}. The formulation for the logarithm of the proton decay half-life can be expressed as follows:
\begin{equation}
\label{pudl}
\log T_{1 / 2}=a \chi^{\prime}+b \rho^{\prime}+d \, \ell(\ell+1) / \rho^{\prime}+c,
\end{equation}
where $a, b, c$, and $d$ are constants, $\chi^{\prime}=A^{1 / 2}  Z_{d} Q_{p}^{-1 / 2}$, $\rho^{\prime}=\sqrt{A  Z_{d}\left(A_{d}^{1 / 3}+1\right)}$, and  $A=A_{d}  /\left(A_{d}+1\right)$.

The constants $a, b, c$, and $d$ are established through the process of fitting them to existing experimental data. It's particularly intriguing to explore how well this formula aligns with our computed values for proton decay half-lives using the Densiov Potential. To investigate this, we conducted a comprehensive search to determine the optimal values for the four free parameters ($a, b, c$, and $d$) in the context of Densiov Potential calculations, as well as with respect to the most recent experimental data. In Figure~\ref{Fig4}, we present the relationship between the quantity $\log T_{1/2}^{p} -e$ and $\chi'$. Here, $e$ is defined as $e = b,\rho' + d ,\ell(\ell + 1)/\rho' + c$. Figure~\ref{Fig4} illustrates this relationship using computed values for proton decay half-lives using the Denisov potential, denoted as $T_{1/2}^{\text{p}}$. It's evident from Figure~\ref{Fig4} that the quantity $\log T_{1/2}^{\text{p}} -\left[b,\rho' + d ,\ell(\ell + 1)/\rho' + c \right]$ exhibits a linear correlation with $\chi'$ in accordance with Eq.~(\ref{pudl}).

\section{\label{sec:level4}SUMMARY AND CONCLUSIONS}

An investigation of the proton emission and alpha decay half-life times of different nuclei has been done, calculations were done without taking temperature once as well as taking it into account. The calculated half-life times agree relatively well with the experimental data. We also see that for the isotopes $^{109}\textrm{I}$, $^{160}\textrm{Re}$, $^{165}\textrm{Ir}^m$, $^{185}\textrm{Bi}$, $^{161}\textrm{Re}$ and $^{112}\textrm{Cs}$, the main decay mode is proton emission. These isotopes are found close to the proton-drip line. In this study a new set of fitting parameters are presented for the universal decay law (UDL) on the proton decay half-life times, using the Data found through the Denisov potential. A universal curve for both proton emission, as well as alpha decay, has been investigated, in which a correlation is present between the logarithm of the experimental half-lives ($\log_{10} T_{1/2}^{exp.}$) vs the negative of the logarithm of the penetration probability($-\log_{10} P$). We notice that the points lying in the universal curve for alpha decay are a bit more scattered than those in the proton radioactivity case. This difference may be due to using a smaller sample for the alpha decay curve, it could also be due to the uncertainties in the measurements of different parameters like spin
party assignments, decay energies as well as half-life times.

\bibliography{ref}

\clearpage
\newpage
\setlength{\LTcapwidth}{\textwidth}
\begin{longtable*}{@{\extracolsep{\fill}}*{13}{c}}
\caption{Calculation of proton decay half-lives of various neutron deficient nuclei. The superscript $m$ in the different nuclei denotes the isomeric states. The experimental values of $Q_{p}$ are taken from \cite{wang2017ame2016}.}
\label{tab:table1} \\

\hline \hline

  Parent & $Q_{p}$(MeV) & $\ell_{min}$ & Temp(MeV)&  \multicolumn{6}{c}{\hspace{3cm}$\log_{10}T_{1/2}^{p}$ ($T_{1/2}$ in s)}\\ \cline{5-13} \rule{0pt}{4ex}
  &    &     & &\multicolumn{2}{c}{Bass73}& \multicolumn{2}{c}{CW76} & \multicolumn{2}{c}{Denisov} & \multicolumn{2}{c}{Guo2013} & Expt.  \\ \cline{5-6} \cline{7-8} \cline{9-10} \cline{11-12}
  &     &     & &T-IND & T-DEP & T-IND & T-DEP & T-IND & T-DEP & T-IND & T-DEP & \\

 \hline\noalign{\smallskip}
 \endfirsthead

 \noalign{\gdef\Continued{}\gdef\ContTable{}}
   \hline
\endlastfoot

 \multicolumn{8}{c}%
{{ \tablename\ \thetable{} (\textit{continued.}})} \\
\hline \hline

  Parent & $Q_{p}$(MeV) & $\ell_{min}$ & Temp(MeV)&  \multicolumn{6}{c}{\hspace{3cm}$\log_{10}T_{1/2}^{p}$ ($T_{1/2}$ in s)}\\ \cline{5-12} \rule{0pt}{4ex}
  &    &     & &\multicolumn{2}{c}{Bass73}& \multicolumn{2}{c}{CW76} & \multicolumn{2}{c}{Denisov} & \multicolumn{2}{c}{Guo2013} & Expt.  \\ \cline{5-6} \cline{7-8} \cline{9-10} \cline{11-12}
  &     &     & &T-IND & T-DEP & T-IND & T-DEP & T-IND & T-DEP & T-IND & T-DEP & \\

  \hline
\endhead

\hline
\endfoot

\hline \hline
\endlastfoot

%

 ${}^{109}\textrm{I}$ &  $0.820\pm{0.004}$ & 2 & 0.411& -3.031 &-2.914 & -2.125 &-2.008 & -4.304  &-4.188 & -3.767 & -3.616   & $-4.032_{-0.004}^{+0.004}$ \\[1.5mm]


 ${}^{112}\textrm{Cs}$ &  $0.816\pm{0.004}$ & 2 & 0.404 & -2.115 &-2.000 &-1.220 &-1.102 & -3.436  &-3.318 &  -2.876 & -2.725  &$-3.310_{-0.007}^{+0.007}$ \\[1.5mm]


 ${}^{157}\textrm{Ta}$ &  $0.935\pm{0.010}$ & 0 & 0.357 & 1.232 & 1.322& 2.028 &2.118 & -0.257  & -0.167 &    0.438 & 0.557    &$-0.527_{-0.017}^{+0.017}$ \\[1.5mm]


 ${}^{160}\textrm{Re}$ &  $1.267\pm{0.007}$ & 2 & 0.406 & -1.718 & -1.628& -0.953 &-0.863 & -3.462  & -3.372 &   -2.675 & -2.548    &$-3.045_{-0.048}^{+0.079}$ \\[1.5mm]


 ${}^{161}\textrm{Re}$ &  $1.197\pm{0.005}$ & 0 & 0.394 & -1.934 &-1.842 &-1.143 &-1.051 & -3.438  & -3.346 &   -2.736 & -2.612   &$-3.357_{-0.010}^{+0.010}$ \\[1.5mm]


 ${}^{161}\textrm{Re}^m$ &  $1.321\pm{0.005}$ & 5 & 0.413 & 1.222  & 1.311& 1.742 &1.831 & -1.554   & -1.465 &   -0.418 & -0.279    &$-0.678_{-0.009}^{+0.009}$ \\[1.5mm]


 ${}^{165}\textrm{Ir}^m$ &  $1.727\pm{0.070}$ & 5 & 0.461 & -1.931  & -1.853& -1.411 &-1.333 & -4.697  & -4.618 &   -3.561 & -3.428       &$-3.462_{-0.087}^{+0.087}$ \\[1.5mm]


 ${}^{166}\textrm{Ir}$ &  $1.152\pm{0.008}$ & 2 & 0.381 & 0.296 &0.390 & 1.043 &1.136 & -1.508  & -1.415  &   -0.688 & -0.560      &$-0.824_{-0.091}^{+0.091}$ \\[1.5mm]


 ${}^{166}\textrm{Ir}^m$ &  $1.324\pm{0.010}$ & 5 & 0.406 &1.795 & 1.884& 2.303 &2.391 & -1.010  & -0.922 &   0.147 & 0.285      &$-0.076_{-0.026}^{+0.026}$ \\[1.5mm]


 ${}^{167}\textrm{Ir}$ & $1.070\pm{0.004}$ & 0 &0.376 & 0.427 &0.524 & 1.202 &1.298 & -1.142  & 1.046 &   -0.406 & -0.279            &$-0.959_{-0.025}^{+0.025}$ \\[1.5mm]


%
%
%
%

  ${}^{167}\textrm{Ir}^m$ &  $1.245\pm{0.005}$ & 5 &0.394& 2.602 & 2.692& 3.107 &3.198 & -0.210   & -0.119 &    0.954 & 1.092  &$0.875_{-0.009}^{+0.009}$ \\[1.5mm]


 ${}^{170}\textrm{Au}$ &  $1.472\pm{0.012}$ & 2 &0.422 & -2.767  & -2.684& -2.024 &-1.941 & -4.577  & -4.493 &    -3.754 & -3.633   &$-3.487_{-0.060}^{+0.075}$ \\[1.5mm]

 ${}^{170}\textrm{Au}^m$ &  $1.757\pm{0.018}$ & 5 &0.458 & -1.687 &-1.610 & -1.177 &-1.100 & -4.478  & -4.401 &  -3.323 & -3.191   &$-2.971_{-0.028}^{+0.035}$ \\[1.5mm]

 ${}^{171}\textrm{Au}^m$ &  $1.707\pm{0.016}$ & 5 &0.450 & -1.416 &-1.339 & -0.908 &-0.830 & -4.210  & -4.132 &   -3.051 & -2.919      &$-2.585_{-0.013}^{+0.013}$ \\[1.5mm]

 ${}^{177}\textrm{Tl}$ &  $1.155\pm{0.019}$ & 0 & 0.369 & 0.478 & 0.570& 1.231 &1.323 & -1.175  & -1.084 &  -0.396 & -0.273    &$-1.176_{-0.121}^{+0.121}$ \\[1.5mm]


 ${}^{177}\textrm{Tl}^m$ &  $1.962\pm{0.030}$ & 5 &0.472 & -2.858 &-2.786 & -2.353 &-2.282 & -5.651   &-5.579 &  -4.482 & -4.354      &$-3.346_{-0.076}^{+0.076}$ \\[1.5mm]


 ${}^{185}\textrm{Bi}$ &  $1.527\pm{0.081}$ & 0 & 0.416 & -3.697  &-3.620 & -2.950 &-2.872 & -5.345 & -5.269 &   -4.563 & -4.451     &$-4.191_{-0.030}^{+0.030}$ \\[1.5mm]


${}^{113}\textrm{Cs}$ &  $0.973\pm{0.002}$ & 2  & 0.438 &-4.493&-4.493  & -3.596&-3.597&	-5.797& -5.269 &   -5.241 & -5.205    &$-4.752_{-0.010}^{+0.010}$  \\[1.5mm]

${}^{117}\textrm{La}$ &  $0.820\pm{0.003}$ & 2 &0.395 &-1.547&-1.430 &	-0.668&-0.551	&-2.918& -5.797 &   -2.331  & -2.181    &$-1.602_{-0.048}^{+0.048}$  \\[1.5mm]

${}^{121}\textrm{Pr}$ &  $0.890\pm{0.010}$ & 2 &0.402 &-1.935& -1.822&	-1.067&-0.955 &	-3.346	& -2.802 &  -2.738 & -2.592    &$-2.000_{-0.130}^{+0.261}$  \\[1.5mm]
${}^{130}\textrm{Eu}$ &  $1.028\pm{0.015}$ & 2 &0.413 &-2.437&-2.333 &	-1.594&-1.490	&-3.929& -3.324 &  -3.278 & -3.139     &$-3.046_{-0.140}^{+0.236}$  \\[1.5mm]
${}^{131}\textrm{Eu}$ &  $0.947\pm{0.005}$ & 2 &0.396 &-1.417&-1.309 &	-0.577&-0.470 &	-2.919	& -3.825 &  -2.261 & -2.120    &$-1.699_{-0.046}^{+0.046}$  \\[1.5mm]

${}^{135}\textrm{Tb}$ &  $1.188\pm{0.007}$ & 3 &0.431&-2.826&-2.729 &	-2.034&-1.937	&-4.621& -2.812 &   -3.860 & -3.723    &$-3.027_{-0.102}^{+0.152}$  \\[1.5mm]

${}^{140}\textrm{Ho}$ &  $1.094\pm{0.010}$ & 3 &0.408&-1.026&-0.927 &	-0.252&-0.153	&-2.876& -4.525 &   -2.085 & -1.947     &$-2.222_{-0.217}^{+0.217}$  \\[1.5mm]
${}^{141}\textrm{Ho}$ &  $1.177\pm{0.007}$ & 3 &0.420&-2.143&-2.048 &	-1.368&-1.272 &	-3.983& -2.776 &   -3.194 & -3.059    &$-2.387_{-0.011}^{+0.011}$  \\[1.5mm]

${}^{141}\textrm{Ho}^m$ &  $1.243\pm{0.014}$ & 0 &0.431&-4.863&-4.771 &	-4.023&-3.931 &	-6.162& -3.887 &   -5.564 & -5.439   &$-5.137_{-0.018}^{+0.018}$  \\[1.5mm]

${}^{145}\textrm{Tm}$ &  $1.736\pm{0.007}$ & 5 &0.496&-3.712&-3.632 &	-3.144&-3.064 &	-6.361& -6.070 &   -5.314 & -5.174   &$-5.499_{-0.027}^{+0.027}$  \\[1.5mm]

${}^{147}\textrm{Tm}$ &  $1.059\pm{0.003}$ & 5 &0.391&2.812&2.912 &	3.358&3.458 &	0.097& 0.197 &  1.180 & 1.328    &$0.587_{-0.022}^{+0.022}$   \\[1.5mm]

${}^{147}\textrm{Tm}^m$ &  $1.127\pm{0.007}$ & 2 &0.403&-1.947&-1.851 &	-1.147&-1.051 &	-3.570& -3.475 &   -2.844 & -2.713    &$-3.444_{-0.048}^{+0.048}$  \\[1.5mm]

${}^{150}\textrm{Lu}$ &  $1.270\pm{0.002}$ & 5 &0.421&0.881&0.974 &	1.422&1.515 &	-1.845&  -1.753 &   -0.754 & -0.610      &$-1.197_{-0.029}^{+0.029}$  \\[1.5mm]

${}^{151}\textrm{Lu}$ &  $1.241\pm{0.002}$ & 5 &0.415&1.088&1.181 &	1.629&1.722 &	-1.639& -1.546 &  -0.543 & -0.400    &$-0.896_{-0.011}^{+0.011}$  \\[1.5mm]

${}^{151}\textrm{Lu}^m$ &  $1.319\pm{0.010}$ & 2 &0.427&-3.521&-3.432 &	-2.729&-2.640 &	-5.167& -5.078 &   -4.429  & -4.303    &$-4.796_{-0.027}^{+0.027}$  \\[1.5mm]

${}^{155}\textrm{Ta}$ &  $1.453\pm{0.015}$ & 5 &0.440&-0.573&-0.487 &	-0.037&0.049 &	-3.309& -3.223 &   -2.203 & -2.064   &$-2.538_{-0.165}^{+0.225}$  \\[1.5mm]

${}^{156}\textrm{Ta}$ &  $1.020\pm{0.004}$ & 2 &0.373&0.923&1.024 &	1.694&1.794 &	-0.806& -0.706 &   -0.026 & 0.107     &$-0.826_{-0.016}^{+0.016}$  \\[1.5mm]

${}^{156}\textrm{Ta}^m$ &  $1.122\pm{0.008}$ & 5 &0.389&3.202&3.299 &	3.727&3.824 &	0.434& 0.531 &   1.557 & 1.701  &$0.933_{-0.005}^{+0.005}$   \\[1.5mm]

${}^{159}\textrm{Re}^m$ &  $1.600\pm{0.050}$ & 5 &0.454&-1.386&-1.304 &	-0.858&-0.776 &	-4.138& -4.056 &   -3.020 & -2.883    &$-4.665_{-0.087}^{+0.087}$  \\[1.5mm]

${}^{176}\textrm{Tl}$ &  $1.265\pm{0.018}$ & 0 &0.386&-0.865&-0.777 &	-0.109&-0.021 &	-2.504& -2.416 &  -1.734 & 1.613     &$-2.284_{-0.012}^{+0.025}$  \\[1.5mm]

\end{longtable*}


\clearpage
\newpage
\setlength{\LTcapwidth}{\textwidth}
\begin{longtable*}{@{\extracolsep{\fill}}*{15}{c}}  
\caption{Calculation of $\alpha$-decay half-lives of various neutron deficient nuclei. The superscript $m$ in the different nuclei denotes the isomeric states. The experimental values of $Q_{\alpha}$ are taken from \cite{wang2017ame2016}.}
\label{tab:table2} \\

\hline \hline

   Parent & $Q_{\alpha}$(MeV) & $\ell_{min}$ & Temp(MeV)&  \multicolumn{11}{c}{\hspace{3cm}$\log_{10}T_{1/2}^{\alpha}$ ($T_{1/2}$ in s)}\\ \cline{5-15} \rule{0pt}{4ex}
  &  &        & &\multicolumn{2}{c}{Bass73}& \multicolumn{2}{c}{CW76} & \multicolumn{2}{c}{Denisov} &\multicolumn{2}{c}{Guo2013}& \multicolumn{2}{c}{gp77} & Expt.  \\ \cline{5-6} \cline{7-8} \cline{9-10} \cline{11-12} \cline{13-14}
  &  &        & &T-IND & T-DEP & T-IND & T-DEP & T-IND & T-DEP & T-IND & T-DEP &T-IND & T-DEP& \\

 \hline\noalign{\smallskip}
 \endfirsthead

 \noalign{\gdef\Continued{}\gdef\ContTable{}}
   \hline
\endlastfoot

 \multicolumn{8}{c}%
{{ \tablename\ \thetable{} (\textit{continued.}})} \\
\hline \hline

   Parent & $Q_{\alpha}$(MeV) & $\ell_{min}$ & Temp(MeV)&  \multicolumn{11}{c}{\hspace{3cm}$\log_{10}T_{1/2}^{\alpha}$ ($T_{1/2}$ in s)}\\ \cline{5-15} \rule{0pt}{4ex}
  &  &        & &\multicolumn{2}{c}{Bass73}& \multicolumn{2}{c}{CW76} & \multicolumn{2}{c}{Denisov} &\multicolumn{2}{c}{Guo2013}& \multicolumn{2}{c}{gp77} & Expt.  \\ \cline{5-6} \cline{7-8} \cline{9-10} \cline{11-12} \cline{13-14}
  &  &        & &T-IND & T-DEP & T-IND & T-DEP & T-IND & T-DEP & T-IND & T-DEP &T-IND & T-DEP& \\

  \hline
\endhead

\hline
\endfoot

\hline \hline
\endlastfoot

%


 ${}^{109}\textrm{I}$ &  $3.918\pm{0.021}$ & 2 & 0.839 & -1.655 &-0.836& -1.459 &-0.652 & -2.327  &-1.516 & -2.696 & -1.736 & -0.498 & 0.339 & $-0.179_{-0.004}^{+0.004}$ \\[1.5mm]


 ${}^{112}\textrm{Cs}$ &  $3.934\pm{0.059}$ & 0 & 0.829 & -0.922 &-0.098 & -0.744 &0.070 & -1.629  &-0.812 & -1.979 & -1.014 & 0.227 & 1.075  &$-0.725_{-0.027}^{+0.027}$ \\[1.5mm]


 ${}^{157}\textrm{Ta}$ &  $6.355\pm{0.006}$ & 0 & 0.877 & -3.087 &-2.48 & -3.257 &-2.661 & -4.409 &-3.810 & -4.514  &-3.743  &-2.067   & -1.424 &$-1.981_{-0.017}^{+0.017}$ \\[1.5mm]


 ${}^{160}\textrm{Re}$ &  $6.698\pm{0.004}$ & 2 & 0.891 & -3.141 &-2.542 & -3.385 &-2.800 & -4.568 &-3.979 & -4.655  &-3.888  &-2.105  & -1.470 & $-2.040_{-0.048}^{+0.079}$ \\[1.5mm]


 ${}^{161}\textrm{Re}$ &  $6.328\pm{0.007}$ & 0 & 0.864 & -2.101 &-1.489 & -2.335 &-1.736 & -3.517  &-2.914 & -3.598  &-2.823  & -1.084  & -0.437 &$-1.503_{-0.010}^{+0.010}$ \\[1.5mm]


 ${}^{161}\textrm{Re}^m$ &  $6.452\pm{0.007}$ & 0 & 0.872 &-2.573 &-1.967 & -2.796 &-2.203 & -3.977 &-3.380 & -4.058  &-3.289  & -1.559  & -0.919   &$-1.801_{-0.009}^{+0.009}$ \\[1.5mm]


 ${}^{165}\textrm{Ir}^m$ &  $7.003\pm{0.070}$ & 0 & 0.896 & -3.691 &-3.11 & -3.925 & -3.358& -5.130 &-4.560 &-5.190  &-4.441  & -2.694  & -2.078 &$-2.637_{-0.087}^{+0.087}$ \\[1.5mm]


 ${}^{166}\textrm{Ir}$ &  $6.722\pm{0.006}$ & 0 & 0.876 & -2.729 &-2.139 & -2.988 &-2.411 & -4.199  &-3.619 &-4.252  &-3.497  & -1.728& -1.103 &$-1.947_{-0.091}^{+0.091}$ \\[1.5mm]


 ${}^{166}\textrm{Ir}^m$ &  $6.894\pm{0.008}$ & 0 & 0.887 & -3.340 &-2.758 & -3.583 & -3.015& -4.792  &-4.220 &-4.846  &-4.098  & -2.343 & -1.726 &$-1.813_{-0.026}^{+0.026}$ \\[1.5mm]


 ${}^{167}\textrm{Ir}$ &  $6.505\pm{0.003}$ & 0 &0.860 & -1.948 &-1.350 & -2.226  &-1.642 & -3.442  &-2.854 &-3.489  & -2.728  & -0.943  & -0.311&$-1.135_{-0.025}^{+0.025}$ \\[1.5mm]

%
%
%
%


 ${}^{167}\textrm{Ir}^m$ &  $6.680\pm{0.003}$ & 0 &0.871& -2.601  &-2.012 & -2.863 &-2.288 & -4.077  &-3.498 & -4.125  &-3.371  & -1.600 & -0.977 &$-1.426_{-0.009}^{+0.009}$ \\[1.5mm]

  ${}^{170}\textrm{Au}$ &  $7.177\pm{0.015}$ & 0 &0.893 & -3.484 &-2.913 & -3.763 &-3.206 & -5.000  &-4.438 &-5.030  &-4.290  & -2.497  & -1.891  &$-2.579_{-0.060}^{+0.075}$ \\[1.5mm]

 ${}^{170}\textrm{Au}^m$ &  $7.462\pm{0.020}$ & 0 &0.910 & -4.415 &-3.859 & -4.667 &-4.122 & -5.900  &-5.352 & -5.933  &-5.204  & -3.434  & -2.841  &$-2.831_{-0.028}^{+0.035}$ \\[1.5mm]

 ${}^{171}\textrm{Au}^m$ &  $7.344\pm{0.017}$ & 0 & 0.901 & -4.060 &-3.500 & -4.322 &-3.776 & -5.560   &-5.010 & -5.586  &-4.857  & -3.079   & -2.484  &$-2.761_{-0.013}^{+0.013}$ \\[1.5mm]


 ${}^{177}\textrm{Tl}$ &  $7.067\pm{0.007}$ & 0 & 0.869 & -2.355 &-1.787 & -2.703 &-2.149 & -3.978 &-3.421 & -3.969  &-3.234  &-1.373  & -0.771  &$-1.608_{-0.121}^{+0.121}$ \\[1.5mm]


 ${}^{177}\textrm{Tl}^m$ &  $7.874\pm{0.024}$ & 0 &0.916 & -4.980 &-4.448 & -5.253 &-4.733 & -6.519 &-6.000 & -6.515  &-5.809  &-4.019   & -3.451  &$-3.328_{-0.076}^{+0.076}$ \\[1.5mm]


 ${}^{185}\textrm{Bi}$ &  $8.138\pm{0.081}$ & 0 & 0.910 & -5.101 &-4.586 & -5.406  & -4.907& -6.709   &-6.206 & -6.662  &-5.977   & -4.154   & -3.607 & $-3.327_{-0.030}^{+0.030}$ \\[1.5mm]

\end{longtable*}

\begin{figure*}[htp!]
\centering
\label{Fig4}
\includegraphics[width=8.6cm]{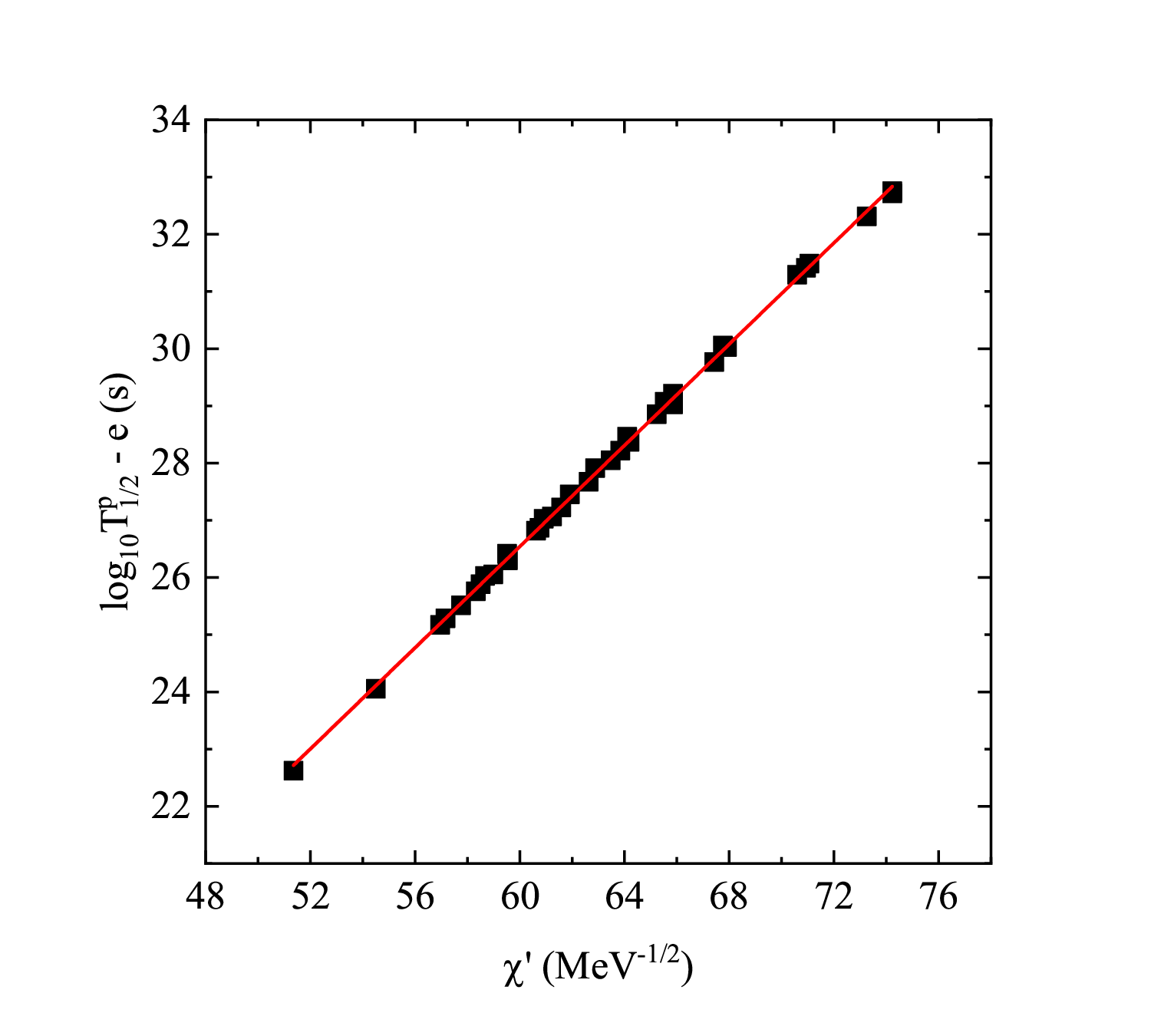}
 \caption{\label{Fig4} UDL plots for the proton decay half-lives illustrating the variation of the quantity $\log_{10} T_{1/2}^p - e$ versus $\chi'$, where $e$ is given by $e=b\hspace{0.1cm}p+d\hspace{0.1cm}l(l+1)/p +c$. Dots are the calculated values of the proton decay half-lives using the Denisov potential.}
\end{figure*}

\end{document}